\documentclass[
 twocolumn,
 superscriptaddress,
 preprintnumbers,
 nofootinbib,
 amsmath,amssymb,
 aps,
 prd,
]{revtex4}
\usepackage{psfrag,slashed,cancel,array,graphicx,hyperref}
\usepackage{subfigure}
\usepackage[utf8]{inputenc}
\usepackage{mathtools}
\usepackage{slashed}
\usepackage{mathrsfs}
\usepackage{multirow}
\usepackage{braket}
\usepackage{float}
\usepackage{booktabs} 
\usepackage[normalem]{ulem}
\hypersetup{pdftitle={},pdfcreator={},linkcolor=[rgb]{0.15,0.35,0.75},colorlinks=true,citecolor=[rgb]{0.675,0,0.2},urlcolor=[rgb]{0.15,0.35,0.65}}

\usepackage{comment}
\usepackage{color} 
\usepackage{graphics}

\newcommand{\asr}{\left(\frac{\alpha_s}{4\pi}\right)}
\newcommand{\ep}{\epsilon}

\newcommand{\msbar}{\overline{\text{MS}}}
\newcommand{\sumQftwo}{\sum_{q'} e_{q'}^2}
\newcommand{\sumLftwo}{\sum_{l} e_{l}^2}
\newcommand{\sumQffour}{\sum_{q'} e_{q'}^4}
\newcommand{\sumLffour}{\sum_{l} e_{l}^4}

\def\asr{\bigg( \frac{\alpha_s}{4 \pi} \bigg)}
\def\aer{\bigg( \frac{\alpha}{4 \pi} \bigg)}

\allowdisplaybreaks

\begin{document}


\title{Renormalization of axial anomaly in SU(N)$\times$U(1)}

\author{Tanmoy Pati}
\email{tanmoy.pati@niser.ac.in}
\affiliation{School of Physical Sciences, National Institute of Science Education and Research, 752050 Jatni, India}
\affiliation{Homi Bhabha National Institute, Training School Complex, Anushakti Nagar, Mumbai 400094, India}

\author{Narayan Rana}
\email{narayan.rana@niser.ac.in}
\affiliation{School of Physical Sciences, National Institute of Science Education and Research, 752050 Jatni, India}
\affiliation{Homi Bhabha National Institute, Training School Complex, Anushakti Nagar, Mumbai 400094, India}

\date{\today}

\begin{abstract}
Defining $\gamma_5$ within dimensional regularization remains a fundamental challenge. Larin's prescription addresses this by introducing additional renormalization constants to restore standard and chiral Ward identities. While these constants are known up to four loops in pure quantum chromodynamics, current precision Standard Model phenomenology requires extending these corrections to mixed gauge sectors. In this article, we propose a novel technique utilizing form factors and the universality of infrared divergences to compute these constants. Applying this framework, we present the new three-loop results for the renormalization constants, as well as
the pure-singlet contributions to the quark axial-vector form factor,
for a mixed $SU(N) \times U(1)$ gauge group.
\end{abstract}



\maketitle


\section{Introduction}
Dimensional regularization (DR) \cite{tHooft:1972tcz} serves as the standard regularization framework for higher-order perturbative calculations in Quantum Field Theory. Despite its widespread success, DR faces a fundamental bottleneck regarding the definition of $\gamma_5$, the matrix crucial for describing chirality. Because a fully anticommuting $\gamma_5$ is incompatible with the Dirac algebra in $d \neq 4$ spacetime dimensions, any consistent prescription must satisfy two key criteria: it must maintain anticommutativity in anomaly-free (non-singlet) sectors to preserve standard chiral symmetries, while simultaneously reproducing the all-order Adler-Bell-Jackiw (ABJ) axial anomaly \cite{Adler:1969gk,Bell:1969ts} in anomalous (pure-singlet) configurations.
One of the most practical approaches, introduced in refs.~\cite{tHooft:1972tcz,Breitenlohner:1977hr} and systematically developed by Larin \cite{Larin:1991tj, Larin:1993tq}, abandons anticommutativity by explicitly extending the four-dimensional definition of $\gamma_5$ via the Levi-Civita tensor into $d = 4 - 2\ep$ dimensions. As a result of this, the Ward identities are violated. However, these identities can be restored order by order in perturbation theory by introducing additional renormalization constants alongside the standard ultraviolet (UV) counterterms. For pure Quantum Chromodynamics (QCD), these renormalization constants have already been computed up to four-loop order \cite{Larin:1991tj, Larin:1993tq, Ahmed:2015qpa, Ahmed:2021spj, Chen:2021gxv, Chen:2022lun} via state-of-the-art multi-loop techniques.

To match the sub-percent precision expected from the High-Luminosity LHC and future colliders data, theoretical predictions must achieve comparable accuracy to ensure that comparisons with data remain statistically meaningful. Achieving this theoretical precision requires moving beyond pure QCD corrections to incorporate mixed QCD-electroweak (EW) effects. Consequently, the renormalization constants inherent to Larin’s prescription must be extended to mixed gauge groups.

In this article, we compute these renormalization constants at three loops within a mixed $SU(N) \times U(1)$ gauge framework. To achieve this, we introduce a novel computational technique that extracts the renormalization constants directly from the three-loop quark form factors by exploiting the universality of infrared (IR) divergences. We first validate our method by reproducing the known three-loop pure QCD results, and then apply it to derive the new corrections for the mixed $SU(N) \times U(1)$ gauge group. Additionally, we analyze the impact of a massive abelian gauge boson and present the three-loop anomalous (pure-singlet) contributions to the quark axial-vector form factors within the same mixed gauge setup.
The remainder of this article is organized as follows. We begin by reviewing Larin's prescription for pure-singlet and non-singlet currents and formulating the anomaly equation for mixed gauge groups. We then discuss our novel form-factor-based computational framework. Finally, we present our three-loop results and conclude.

\section{Theory}
\subsection{Non-singlet Axial-vector and Pseudo-scalar current}
\noindent
We consider the axial-vector and pseudo-scalar currents, $J_\mu^{5}$ and $J^{5}$, 
defined for a massless fermion field $\psi$ as:
\begin{equation}
J_\mu^5 = \bar\psi \gamma_\mu \gamma_5 \psi \,, \quad
J^{5} = \bar\psi \gamma_5 \psi .
\end{equation}
We focus first on the non-singlet case. To treat these currents within DR, 
we adopt Larin's prescription \cite{Larin:1993tq} for the axial-vector current, 
where $\gamma_\mu \gamma_5$ is written as:
\begin{equation}
\gamma_\mu \gamma_5 = \frac{1}{2} (\gamma_\mu \gamma_5 - \gamma_5 \gamma_\mu)
 = \frac{i}{3!} \epsilon_{\mu\nu\rho\sigma}  
   \gamma^\nu \gamma^\rho \gamma^\sigma \,.  
\end{equation}
Similarly, for the pseudo-scalar current, we employ the inherently non-anticommuting definition of $\gamma_5$ in DR, originally introduced in refs.~\cite{tHooft:1972tcz, Breitenlohner:1977hr}:
\begin{equation}
\gamma_5 = \frac{i}{4!} \epsilon_{\mu\nu\rho\sigma}
\gamma^\mu \gamma^\nu \gamma^\rho \gamma^\sigma .
\label{eq:g5def}
\end{equation}
Because these definitions explicitly break the anticommutativity of $\gamma_5$ in $d$ dimensions, standard current properties and conventional Ward identities, which otherwise hold in alternative regularization frameworks like Pauli-Villars regularization, are violated.
Consequently, as shown in ref.~\cite{Larin:1993tq}, restoring the standard chiral properties and 
ensuring the renormalization-group invariance of the non-singlet axial current requires additional 
$\msbar$ renormalization $Z_{\scriptscriptstyle \msbar}^{\text{ns}}$
as well as a finite renormalization $Z_5^{\text{ns}}$. 
The relation between the bare and renormalized non-singlet axial-vector and pseudo-scalar currents are thus given by:
\begin{align}
 (J_\mu^5)_R &= Z_{\scriptscriptstyle \msbar}^{\text{ns}} Z_5^{\text{ns}} (J_\mu^5)_B \,, \quad 
 (J^5)_R = Z_{\scriptscriptstyle \msbar}^{\text{p}} Z_5^{\text{p}} (J^5)_B \,.
 \label{eq:nsren} 
\end{align}
These renormalization constants can be determined by demanding that the renormalized vertices, with and without $\gamma_5$, coincide.

\subsection{Pure-singlet Axial-vector Current}
Let us now consider the pure-singlet axial-vector current. In analogy with the non-singlet case, 
the relation between the bare and renormalized singlet current is defined via
\begin{equation}
 (J_\mu^5)_R = Z_{\scriptscriptstyle \msbar}^{\text{s}} Z_5^{\text{s}} (J_\mu^5)_B \,.
 \label{eq:sing}
\end{equation}
While $Z_{\scriptscriptstyle \msbar}^{\text{s}}$ is determined by absorbing the UV divergences, 
the computation of the finite renormalization constant $Z_5^{\text{s}}$ differs from the non-singlet case. 
In the pure-singlet case, because of the non-trivial axial Ward identity, one cannot simply impose the equality of vector and axial-vector vertices.
Instead, $Z_5^{\text{s}}$ can be obtained by restoring the correct form of the axial Ward identity, namely, the all-order axial-anomaly equation.
For a mixed $SU(N) \times U(1)$ gauge group, the operator equation can be formulated in terms of renormalized local composite operators as:
\begin{equation}
\label{eq:anomaly}
 \left(\partial^\mu J_\mu^5\right)_R =  a_s n_F T_F (G\widetilde{G})_R + a_e (F\widetilde{F})_R \,,
\end{equation}
where $a_s = \alpha_s / (4\pi)$ and $a_e = \alpha_e / (4\pi)$,
with $\alpha_s$ and $\alpha$ being the gauge coupling parameters for the $SU(N)$ and $U(1)$ gauge groups, respectively. The composite operators are defined as 
\begin{align}
 G\widetilde{G} &= G_{\mu\nu}^a \widetilde{G}^{a,\mu\nu}
                = \epsilon_{\mu\nu\rho\sigma} G_{a,\mu\nu} G^{a}_{\rho\sigma} ,
 \nonumber\\
 F\widetilde{F} &= F_{\mu\nu} \widetilde{F}^{\mu\nu}
                = \epsilon_{\mu\nu\rho\sigma} F_{\mu\nu} F_{\rho\sigma} .
\end{align}
$G_{\mu\nu}^a$ and $F_{\mu\nu}$
represent the field strength tensors of the respective gauge fields.
We note that, because of the mixed gauge group, the axial-vector current is coupled to both massless
quarks ($q$) and leptons ($l$).

\section{Methodology}
To determine the non-singlet renormalization constants $Z_{\scriptscriptstyle \msbar}^{\text{ns}}$, $Z_5^{\text{ns}}$, $Z_{\scriptscriptstyle \msbar}^{p}$ and $Z_5^{p}$, we compute the three-loop non-singlet
quark form factors i.e. the matrix elements of the vector, axial-vector, scalar and pseudo-scalar currents
between external on-shell quark states. 
The pure-singlet renormalization constant $Z_{\scriptscriptstyle \msbar}^{\text{s}}$ is obtained by evaluating 
the three-loop matrix elements of the pure-singlet axial-vector current.
The finite renormalization constant $Z_5^{\text{s}}$ is determined
from the three-loop matrix elements of the anomaly operator equation, Eq.~\eqref{eq:anomaly}, 
between external quark states, namely $\langle q| \partial^\mu J_\mu^5 | q \rangle$. 
We employ the standard automated framework for multi-loop computations to evaluate the required three-loop form factors.
The relevant Feynman diagrams are generated using \textsc{QGRAF}~\cite{Nogueira:1991ex}. 
Subsequent symbolic manipulation, including the evaluation of Lorentz, Dirac, and gauge group algebra, 
is performed via in-house \textsc{FORM}~\cite{Vermaseren:2000nd, Ruijl:2017dtg} routines. 
The resulting large set of scalar loop integrals is reduced to a minimal basis of Master Integrals (MIs) via Integration-By-Parts (IBP) identities using \textsc{Kira 3}~\cite{Klappert:2020aqs}.
The evaluation of diagrams containing gluon and the massless $U(1)$ gauge field is well-established 
with the MIs been known for some time~\cite{Gehrmann:2010ue}.
In contrast, the evaluation of diagrams with a massive $U(1)$ gauge field introduces significant computational complexity, particularly within the IBP reduction step. The use of Larin's prescription results in longer Dirac traces, which further 
generate more complex seed integrals compare to \cite{Pati:2025xht}.
The MIs corresponding to this case were recently evaluated by us in ref.~\cite{Pati:2025ivg}.
Alongside the three-loop diagrams, the evaluation of one- and two-loop contributions also is required. 
After obtaining the bare form factors, we performed standard UV renormalization, along with the special renormalization, Eq.~\eqref{eq:sing}, required for restoring the chiral properties.

For the non-singlet case, 
an order-by-order comparison in the coupling constants $\alpha_s$, $\alpha$, and the dimensional regulator $\epsilon$ between the UV renormalized axial-vector and vector form factors yields the results for $Z_{\scriptscriptstyle \msbar}^{\text{ns}}$ and $Z_5^{\text{ns}}$ at three loops. 
Similarly, a direct comparison between the pseudo-scalar and scalar form factors provides the corresponding three-loop expressions for $Z_{\scriptscriptstyle \msbar}^{\text{p}}$ and $Z_5^{\text{p}}$.
For the pure-singlet case, $Z_{\scriptscriptstyle \msbar}^{\text{s}}$ is obtained by demanding the UV finiteness of the pure-singlet quark axial-vector form factors. 
Computing $Z_5^{\text{s}}$ requires evaluating the matrix elements of the anomaly equation i.e. $\langle q| \partial^\mu J_\mu^5 | q \rangle$, $\langle q| G \widetilde{G} | q \rangle$ and $\langle q| F \widetilde{F} | q \rangle$. 
At the two-loop level (for $\mathcal{O}(\alpha_s^2)$ and $\mathcal{O}(\alpha^2)$), 
both the left-hand and right-hand sides of Eq.~\eqref{eq:anomaly} are free of IR divergences, 
allowing $Z_5^{\text{s}}$ to be extracted immediately.
Beyond two loops, the presence of IR poles in the form factors prohibits a direct algebraic 
comparison between the two matrix elements in Eq.~\eqref{eq:anomaly}. 
To overcome this issue, we exploit the universal structure of IR singularities in gauge theories. 
By factoring out these universal IR poles, we define a finite remainder for the form factors.
In the following sections, we describe the implementation of UV renormalization and IR subtraction frameworks to construct these finite remainders and finally extract the three-loop finite singlet renormalization constant $Z_5^{\text{s}}$.

\subsection{UV renormalization}
We begin by performing standard UV renormalization of the fields and parameters.
In the case of $SU(N) \times U(1)$ with a massless $U(1)$ gauge field, it is sufficient to renormalize the coupling constants, $\alpha_s$ and $\alpha$, by multiplying them by their respective renormalization constants, $Z_{\alpha_s}$ and $Z_{\alpha}$. If the $U(1)$ gauge field is massive, an additional renormalization of the quark wave function is required. 
Following ref.~\cite{Billis:2019evv}, we adopt the definitions for $Z_{\alpha_s}$ and $Z_{\alpha}$, 
and define the corresponding $\beta$-functions as
\begin{align}
 \beta_{a_s} &= - 2 \alpha_s \sum_{i,j=0}^\infty \beta_{ij}  \asr^{i+1} \aer^j \,, 
 \\
 \beta_{a_e} &= - 2 \alpha_e \sum_{i,j=0}^\infty \beta_{ij}^e \asr^i \aer^{j+1} \,.
\end{align}
In addition to the coupling constant renormalization, special renormalization steps,
as given in Eqs.~(\ref{eq:nsren},\ref{eq:sing}), are required to restore chiral properties. Because our primary goal is to derive these renormalization constants, we retain their symbolic forms throughout the intermediate steps.
As discussed above, for the non-singlet case, the renormalization constants can be obtained via a direct comparison between the vector and axial-vector or scalar and pseudo-scalar form factors. For the singlet case, however, the renormalization constants must be determined using Eq.~\eqref{eq:anomaly}, which holds exclusively for renormalized operators. 
The renormalization of $\partial^\mu J_\mu^5$ is purely multiplicative and depends solely on its bare counterpart, $(\partial^\mu J_\mu^5)_B$. The renormalization of the remaining two operators, $G \widetilde{G}$ and $F \widetilde{F}$, 
mixes with $\partial^\mu J_\mu^5$. The renormalization of all three operators
can be arranged in the following matrix form 
\begin{equation*}
\begin{pmatrix}
 ( G \widetilde{G} )_R \\ ( F \widetilde{F} )_R \\ ( \partial^\mu J_\mu^5 )_R
\end{pmatrix}=\begin{pmatrix}
  Z_{GG} & 0 & Z_{GJ} \\ 0 & Z_{FF} & Z_{FJ} \\ 0 & 0 & Z_{JJ}
\end{pmatrix}\begin{pmatrix}
 ( G \widetilde{G} )_B \\ ( F \widetilde{F} )_B \\ ( \partial^\mu J_\mu^5 )_B
\end{pmatrix}
\end{equation*}
where the elements of this mixing matrix admit perturbative expansion in $\alpha_s$ and $\alpha$ within the context of the mixed gauge group.

\subsection{Extracting the renormalization constants}
The IR structure of a generic scattering amplitude in an $SU(N)$ gauge theory is universal. 
This universal behavior was first systematically presented for one- and two-loop QCD amplitudes 
by Catani~\cite{Catani:1998bh} and Sterman~\cite{Sterman:2002qn} through the introduction of subtraction operators. 
Later, the factorization of the single pole of form factors in terms of soft and collinear 
anomalous dimensions was predicted up to two loops~\cite{Ravindran:2004mb} 
and later confirmed at three loops~\cite{Moch:2005tm} in pure QCD. 
The IR properties of three-loop form factors in a mixed $SU(N) \times U(1)$ gauge theory were 
discussed in detail in Ref.~\cite{AH:2019pyp}. Catani's framework has also been generalized beyond two 
loops in Refs.~\cite{Becher:2009cu, Gardi:2009qi}, where the authors introduced 
a multiplicative IR renormalization factor, $Z_{\text{IR}}$,
which contains all IR divergences so that the remaining component 
is finite as $\epsilon \rightarrow 0$. In the case of form factors, this relation is expressed as
$F = Z_{\text{IR}}(\mu) F^{\text{fin}}(\mu)$.
The operator $Z_{\text{IR}}$ is governed by a renormalization group equation (RGE) as 
\begin{equation}
 \frac{d \ln Z_{IR}}{d \ln \mu} = - \Gamma = - \sum_{m+n \ge 1}^\infty \Gamma_{mn} a_s^{m} a_e^{n} 
\end{equation}
Its solution can be expressed in terms of the relevant anomalous dimensions ($\Gamma_{mn}$) and $\beta$-functions. While the RGE in ref.~\cite{Becher:2009cu} depends solely on the strong coupling $\alpha_s$ due to purely QCD corrections, the situation changes when considering a mixed $SU(N) \times U(1)$ gauge group. In this setup, the RG evolution equations of the couplings become mutually dependent, necessitating the solution of a system of coupled first-order differential equations. Consequently, $Z_{\text{IR}}$ depends on both QCD ($\beta_{a_s}$) and QED ($\beta_{a_e}$) $\beta$-functions, which we have defined above. The explicit solution for $Z_{\text{IR}}$ up to three loops in the mixed $SU(N) \times U(1)$ theory is given by
\begin{widetext}
\begin{align}
 \ln Z_{\text{IR}} &= 
 \frac{\alpha_s}{4 \pi} \bigg( 
 \frac{\Gamma_{10}'}{4 \epsilon^2} + \frac{\Gamma_{10}}{2 \epsilon} \bigg)
+\frac{\alpha}{4 \pi} \bigg( 
 \frac{\Gamma_{01}'}{4 \epsilon^2} + \frac{\Gamma_{01}}{2 \epsilon} \bigg)
+ \frac{\alpha_s}{4 \pi} \frac{\alpha}{4 \pi} \bigg( \frac{\Gamma_{11}'}{16 \epsilon^2} + \frac{\Gamma_{11}}{4 \epsilon} \bigg)             
\nonumber\\&
+ \asr^2 \bigg( -\frac{3 \beta_{00} \Gamma_{10}'}{16 \epsilon^3} 
              + \frac{\Gamma_{20}' - 4 \beta_{00} \Gamma_{10}}{16 \epsilon^2}  
              + \frac{\Gamma_{20}}{4 \epsilon} \bigg) 
+ \aer^2 \bigg( -\frac{3 \beta^{e}_{00} \Gamma_{01}'}{16 \epsilon^3} 
              + \frac{\Gamma_{02}' - 4 \beta_{00}^e \Gamma_{01}}{16 \epsilon^2} 
              + \frac{\Gamma_{02}}{4 \epsilon} \bigg)
\nonumber\\&
+ \asr^3 \bigg( \frac{11 \beta_{00}^2 \Gamma_{10}'}{72 \epsilon^4} 
              - \frac{8 \beta_{10} \Gamma_{10}' + 5 \beta_{00} \Gamma_{20}' - 12 \beta_{00}^2 \Gamma_{10}}{72 \epsilon^3} 
              + \frac{\Gamma_{30}' - 6 \beta_{10} \Gamma_{10} - 6 \beta_{00} \Gamma_{20}}{36 \epsilon^2} 
              + \frac{\Gamma_{30}}{6 \epsilon} \bigg)
\nonumber\\&
+ \asr^2 \aer \bigg( -\frac{16 \beta_{01} \Gamma_{10}' +  5 \beta_{00} \Gamma_{11}'}{144 \epsilon^3} 
              + \frac{\Gamma_{21}' - 6 \beta_{01} \Gamma_{10} - 3 \beta_{00} \Gamma_{11}}{36 \epsilon^2} 
              + \frac{\Gamma_{21}}{6 \epsilon} \bigg)
\nonumber\\&
+ \asr \aer^2 \bigg( -\frac{16 \beta_{10}^e \Gamma_{01}' + 5 \beta_{00}^e \Gamma_{11}'}{144 \epsilon^3}  
              + \frac{\Gamma_{12}' - 6 \beta_{10}^e \Gamma_{01} - 3 \beta_{00}^e \Gamma_{11}}{36 \epsilon^2}  
              + \frac{\Gamma_{12}}{6 \epsilon} \bigg)
\nonumber\\&
+ \aer^3 \bigg( \frac{11 (\beta_{00}^e )^2 \Gamma_{01}'}{72 \epsilon^4} 
              - \frac{8 \beta_{01}^e \Gamma_{01}' + 5 \beta_{00}^e \Gamma_{02}' - 12 (\beta_{00}^e)^2 \Gamma_{01}}{72 \epsilon^3} 
              + \frac{\Gamma_{03}' - 6 \beta_{01}^e \Gamma_{01} - 6 \beta_{00}^e \Gamma_{02}}{36 \epsilon^2} 
              + \frac{\Gamma_{03}}{6 \epsilon} \bigg) \,.
\end{align}
\end{widetext}
To calculate the pure-singlet renormalization constants, we first extract the pure-singlet contributions for the operator $\partial_\mu J_\mu^5$ at each order in $\alpha_s$ and $\alpha$ up to three loops. In this step, the renormalization constants $Z_{\msbar}^s$ and $Z_{5}^s$ are kept as symbols. We then perform IR subtraction by multiplying the UV renormalized matrix element by $Z_{\text{IR}}^{-1}$ within this singlet category.
By demanding the universality of IR divergences, the IR subtracted contributions 
must be finite as $\ep \rightarrow 0$. So, equating the coefficients of each pole in $\ep$ allows us to determine $Z_{\msbar}^s$ at three loops. 
The resulting finite remainder is given by
\begin{equation}
    \langle q | \partial_\mu J_\mu^5 | q \rangle^{\text{f}} = \left. Z_{\text{IR}}^{-1} \, \langle q | \partial_\mu J_\mu^5 | q \rangle_{\text{R}} \right|_{\epsilon \rightarrow 0} \,.
\end{equation}
Next, following the procedure discussed in the previous section,
we perform UV renormalization for the contributions from the other two 
operators, $G\widetilde{G}$ and $F\widetilde{F}$.
After applying appropriate IR subtraction to these terms, we obtain the finite remainders
\begin{align}
 \langle q | G\widetilde{G} | q \rangle^{\text{f}} &= \left. Z_{\text{IR}}^{-1} \, \langle q | G\widetilde{G} | q \rangle_{\text{R}} \right|_{\epsilon \rightarrow 0} \,, 
 \\
 \langle q | F\widetilde{F} | q \rangle^{\text{f}} &= \left. Z_{\text{IR}}^{-1} \, \langle q | F\widetilde{F} | q \rangle_{\text{R}} \right|_{\epsilon \rightarrow 0} \,.
\end{align}
Finally, the axial anomaly equation can be expressed directly in terms of these finite remainders as
\begin{equation}
    \langle q | \partial^\mu J_\mu^5 | q \rangle^{\text{f}} = a_s n_f T_F \langle q | G\widetilde{G} | q \rangle^{\text{f}} + a_e \langle q | F\widetilde{F} | q \rangle^{\text{f}} \,.
\label{eq:anomalyFin}
\end{equation}
We note that at three loops in pure QCD, singlet diagrams are classified into two categories based on the number of propagators in the internal fermion loop: triangle-type and box-type topologies, which contain three and four internal fermion propagators, respectively. 
The box-type diagrams also yield non-singlet finite contributions proportional to the color factor $C_A C_F n_f T_F$. 
On the other hand, these contributions vanish in mixed QCD-QED and in pure three-loop QED.
While our procedure allows for the evaluation of two color factors ($C_F^2 n_f T_F$ and $C_F n_f^2 T_F^2$) in pure QCD, it enables the calculation of the complete result in the mixed QCD-QED and pure QED framework.

\subsection{Abelianization}
The abelianization procedure uses certain transformation rules that relate color and charge factors of the relevant Feynman diagrams, to obtain QED and mixed QCD-QED results 
directly from pure QCD expressions. 
While this approach has been extensively applied at the two-loop level \cite{deFlorian:2015ujt,deFlorian:2016gvk,deFlorian:2018wcj,AH:2019xds}, its limitations become apparent at three loops.
As shown in ref.~\cite{AH:2019pyp} through explicit computations of the form factors, 
there exists definite transformation rules for the color factors $C_F^3, C_A C_F^2, C_A^2 C_F$ and $C_F n_f^2 T_F^2$ in pure QCD.
However, the procedure fails for the color factor $C_F^2 n_f T_F$ which originates from topologies with a single internal fermion loop.
For this specific case, 
there is no one-to-one mapping from pure QCD to either pure QED or to mixed QCD-QED.
This is due to the fact that while gluons do not see the flavor charge of the quarks, photons do.
For completeness, we present the transformation rules of ref.~\cite{AH:2019pyp} for three loops in Table~\ref{tab:abel}.
\begin{table}[h]
\centering
\renewcommand{\arraystretch}{1.5}
\begin{tabular}{|c||c|c|c|}
\hline
 $a_s^3$ & $a_s^2 a_e$ & $a_s a_e^2$ & $a_e^3$ \\
\hline\hline
$C_F^3$ & $3 C_F^2 e_u^2$ & $3 C_F e_u^4$ & $e_u^6$ \\
\hline
$C_A C_F^2$ & $C_A C_F e_u^2$ & $0$ & $0$ \\
\hline
$(a + b)\;C_F^2 n_f T_F$ & 
\begin{tabular}{@{}c@{}} $a \, C_F n_f T_F e_u^2$ \\ $b \, C_F T_F \mathcal{Q}_q$ \end{tabular} & 
\begin{tabular}{@{}c@{}} $a \, C_F e_u^2 \mathcal{Q}_f $ \\ $b \, C_F e_u^2  N \mathcal{Q}_q$ \end{tabular} & 
\begin{tabular}{@{}c@{}} $a \, e_u^4 \mathcal{Q}_f$ \\ $b \, e_u^2 \mathcal{Q}_f^{(2)} $ \end{tabular} \\
\hline
$C_F C_A^2$ & $0$ & $0$ & $0$ \\
\hline
$C_F C_A n_f T_F$ & $0$ & $0$ & $0$ \\
\hline
$C_F n_f^2 T_F^2$ & $0$ & $0$ & $e_u^2 \big(\mathcal{Q}_f\big)^2$ \\
\hline
\end{tabular}
\caption{Abelianization rules at three loops.}
\label{tab:abel}
\end{table}
$C_A=N$ and $C_F = (N^2-1)/2/N$ are the adjoint and fundamental Casimirs of $SU(N)$, respectively.
$T_F = 1/2$.
$n_f$ and $n_l$ are the number of massless quarks and leptons, respectively.
The $\mathcal{Q}$-symbols are defined as
\begin{align}
 \mathcal{Q}_q &=  \sumQftwo   \,, \,
 \mathcal{Q}_q^{(2)} =  \sumQffour \,, \,
 \mathcal{Q}_f^{(2)} = \big( \mathcal{Q}_l^{(2)} + N \mathcal{Q}_q^{(2)} \big)\,,
 \nonumber\\
 \mathcal{Q}_l &= \sumLftwo  \,, \,
 \mathcal{Q}_l^{(2)} = \sumLffour \,, \,
 \mathcal{Q}_f = ( \mathcal{Q}_l + N  \mathcal{Q}_q ) \,.
\end{align}
Two important remarks are in order. First, while the parameters $a$ and $b$ vary across different quantities, such as form factors and renormalization constants, they remain same for any single given object across different perturbative orders. For instance, for the case of $Z_5^{s}$, 
the values of $a$ and $b$ are same at each order in the coupling constant at three loops.
Second, considering the axial-vector current in the full SM, the current features a coupling proportional to the weak isospin of the fermion. If one sums the anomalous contributions from a single generation of quarks and leptons, the anomalous contributions cancel, at the two-loop level, since $\mathcal{Q}_f = 0$.
However, beyond two loops and upon considering mixed corrections, the renormalization constants receive additional contributions that do not purely scale with $\mathcal{Q}_f$. 
These terms are not necessarily zero even for the SM axial-vector current, 
preventing the cancellation.

\section{Results}
All the renormalization constants admit series expansion in $a_s$ and $a_e$ as follows
\begin{align}
 Z_{\scriptscriptstyle \msbar}^{\text{I}} &= \sum_{m,n=0} a_s^m a_e^n Z_{\text{I},\epsilon}^{(m,n)} \,,
 \\
 Z_5^{\text{I}} &= \sum_{m,n=0} a_s^m a_e^n Z_{\text{I},f}^{(m,n)}
\end{align}
I = ns, s and p denote the non-singlet axial-vector, pure-singlet axial-vector and pseudo-scalar contributions, 
respectively.
The pure QCD three-loop results $Z_{\text{ns},\epsilon}^{(3,0)}$, $Z_{\text{s},\epsilon}^{(3,0)}$,
$Z_{\text{p},\epsilon}^{(3,0)}$, $Z_{\text{ns},f}^{(3,0)}$ and $Z_{\text{p},f}^{(3,0)}$
were presented in refs.~\cite{Larin:1991tj,Larin:1993tq}
whereas $Z_{\text{s},f}^{(3,0)}$ has been computed in ref.~\cite{Ahmed:2021spj}.
One can use abelianization rules of ref.~\cite{AH:2019pyp}, as presented in Table~\ref{tab:abel}, 
to obtain partial results for $Z_{\text{I},\epsilon}^{(m,n)}$ and $Z_{\text{I},f}^{(m,n)}$
for $(m,n) = (2,1), (1,2)$ and $(0,3)$. 

Thanks to our novel technique of using the universal IR structure, we obtain the complete results for these renormalization constants through explicit three-loop computations. 
To present these concisely, we provide the values of $a$ and $b$ in Table~\ref{tab:valab}.
Using these values along with the abelianization rules provided in Table~\ref{tab:abel},
one can obtain all renormalization constants in the mixed gauge group from 
the known three-loop QCD results. We also present them explicitly in the supplemental text.

\begin{table}[H]
\centering
\renewcommand{\arraystretch}{1.5}
\begin{tabular}{|c||c|c||}
\hline
  & ~~~a~~~ & ~~~b~~~  \\
\hline\hline
$Z_{ns, \epsilon}^{\rm 3l}$ & $\frac{112}{9} \frac{1}{\ep}$ & $-\frac{16}{3} \frac{1}{\ep}$ \\
\hline
$Z_{ns, f}^{\rm 3l}$ & $-\frac{568}{27}$ & $\frac{148}{9} -\frac{64}{3} \zeta_3$ \\
\hline
$Z_{s, \epsilon}^{\rm 3l}$ & $-\frac{28}{\ep}$ & $\frac{16}{\ep}$ \\
\hline
$Z_{s, f}^{\rm 3l}$ & $-173$ & $196 - 48 \zeta_3$ \\
\hline
$Z_{p, \epsilon}^{\rm 3l}$ & $\frac{6}{\ep^3} + \frac{31}{3 \ep^2} - \frac{253}{9 \ep}$ 
                           & ~$- \frac{4}{\ep^2} + \left( \frac{13}{3} - 16 \zeta_3 \right) \frac{1}{\ep}$~  \\
\hline
$Z_{p, f}^{\rm 3l}$ & $-\frac{2048}{27}$ & $\frac{296}{9} - \frac{128}{3} \zeta_3$ \\
\hline
$\mathcal{F}_{s,f}^{\rm 3l}$ & ~$409-96 \zeta_2-\frac{192}{5}\zeta_2^2$~ & $32\zeta_2$ \\
\hline
\end{tabular}
\caption{The values of $a$ and $b$ at three loops.}
\label{tab:valab}
\end{table}
%

Once we obtain the renormalization constants using the anomaly equation,
we employ them to compute the renormalized, IR-subtracted form factor 
$\langle q | J_\mu^5 | q \rangle^{\text{f}}$, i.e., the pure-singlet contributions to the three-loop corrections for the quark axial-vector form factor within a mixed $SU(N)\times U(1)$ gauge group. The three-loop pure QCD result was obtained in ref.~\cite{Gehrmann:2021ahy}.
While the full expressions of the form factors are presented in the supplemental text, 
we provide the values of $a$ and $b$ for this case also in Table~\ref{tab:valab}. 

As mentioned earlier, we also consider a massive $U(1)$ gauge boson to verify 
the validity of these renormalization constants. As expected, giving a mass to 
the $U(1)$ gauge boson does not affect the UV behavior originating from $\gamma_5$. 
We explicitly verify the Ward identities at two loops for massive $U(1)$ gauge fields.
We also confirm the validity of the renormalization constants at ${\mathcal{O}}(\alpha \alpha_s^2)$.

\section{Conclusion}
The High-Luminosity LHC and future colliders will collect data with exceptional precision,
demanding theoretical predictions to account for mixed QCD-EW effects moving beyond pure QCD corrections. On the other hand, the definition of $\gamma_5$ is a fundamental challenge within DR. While Larin's prescription offers a practical and widely adopted framework,
it requires the introduction of additional renormalization constants alongside standard UV counterterms to restore broken Ward identities. These scheme-dependent constants must be computed order-by-order in perturbation theory.

In this article, we introduce a novel approach to compute these renormalization constants
using massless quark form factors and IR universality.
To demonstrate the validity of our method, we first reproduce 
the known three-loop pure QCD results
for two color factors $C_F^2 n_f T_F$ and $C_F n_f^2 T_F^2$, finding perfect agreement with the literature. We then apply this framework to compute, for the first time, all the renormalization constants for the axial-vector and pseudo-scalar currents at three loops in mixed $SU(N) \times U(1)$ gauge group. 
We also demonstrate the known fact that the abelianization procedure fails beyond two loops,
confirming that mixed QCD-QED results cannot be simply predicted from pure QCD expressions without
explicit computation. 
Further, we consider a massive $U(1)$ gauge theory and check that 
introducing a mass to the $U(1)$ gauge boson does not affect the UV behavior originating 
from $\gamma_5$. The mass-independent nature of these renormalization constants 
allows this work to serve as a crucial first step toward the computation of three-loop mixed QCD-EW corrections. 
Finally, we present the pure-singlet contributions to the three-loop corrections for the quark axial-vector form factor within the mixed $SU(N)\times U(1)$ gauge group.

\subsection{Acknowledgements}
We would like to thank S. Moch and A. Vicini for fruitful discussions.

\begin{widetext}

\appendix

\section{Non-singlet and pure-singlet renormalization constants}
\label{app:res}
In the following, we present the three-loop constants for the axial-vector current. For completeness, we also present the pure QCD results.
%
\begin{align}
{ Z_{ns, \epsilon}^{(3,0)}} &=
 \frac{1}{\ep^2} \bigg\{ 
-\frac{484}{27} C_A^2 C_F
+\frac{352}{27} C_A C_F n_F T_F
-\frac{64}{27} C_F n_F^2 T_F^2 
 \bigg\}
+\frac{1}{\ep} \bigg\{  
\frac{3578}{81} C_A^2 C_F
-\frac{308}{9} C_A C_F^2
-\frac{1664}{81} C_A C_F n_F T_F
\nonumber\\&
+\frac{64}{9} C_F^2 n_F T_F
+\frac{32}{81} C_F n_F^2 T_F^2
\bigg\}\,.
\\
 {Z_{ns, \epsilon}^{(2,1)}} &=
\frac{1}{\ep} \bigg\{  
-\frac{308}{9} C_A C_F e_u^2
+\frac{112}{9} C_F n_F T_F e_u^2
-\frac{16}{3}  C_F T_F \mathcal{Q}_q 
\bigg\}\,.
\\
 {Z_{ns, \epsilon}^{(1,2)}} &=
\frac{1}{\ep} \bigg\{  
\frac{112}{9} C_F e_u^2  \mathcal{Q}_f 
-\frac{16}{3} C_F e_u^2 N \mathcal{Q}_q 
\bigg\}\,.
\\
 {Z_{ns, \epsilon}^{(0,3)}} &=  
\frac{1}{\ep^2} \bigg\{  
-\frac{64}{27} e_u^2 \big( \mathcal{Q}_f \big)^2 
\bigg\}
+ \frac{1}{\ep} \bigg\{ 
\frac{112}{9} e_u^4  \mathcal{Q}_f 
- \frac{16}{3} e_u^2 \mathcal{Q}_f^{(2)} 
+ \frac{32}{81} e_u^2 \big( \mathcal{Q}_f \big)^2
\bigg\}\,.
%
\\
%
{Z_{ns,f}^{(3,0)}} &=
\bigg(
        -\frac{2147}{27}
        +56 \zeta_3
\bigg) C_A^2 C_F
+\bigg(
                \frac{5834}{27}
                -160 \zeta_3
\bigg) C_A C_F^2
+\bigg(\frac{712}{81}
        +\frac{64}{3} \zeta_3
\bigg) C_A C_F n_F T_F
+\bigg(
        -\frac{370}{3}
        +96 \zeta_3
\bigg) C_F^3
\nonumber\\&
+\big(-\frac{124}{27}
        -\frac{64}{3}\zeta_3
\big) C_F^2 n_F T_F
+\frac{208}{81} C_F n_F^2 T_F^2
\,.
%
\\
{ Z_{ns,f}^{(2,1)}} &=
\bigg(
        \frac{5834}{27}
        -160 \zeta_3
\bigg) C_A C_F e_u^2
+\big(
        -370
        +288 \zeta_3
\big) C_F^2 e_u^2
-\frac{568}{27} C_F n_F T_F e_u^2
+ \bigg(
        \frac{148}{9}
        -\frac{64}{3} \zeta_3 
\bigg) C_F T_F \mathcal{Q}_{q} \,.
%
\\
 {Z_{ns,f}^{(1,2)}} &=
\big(
        -370
        +288 \zeta_3
\big) C_F e_u^4
-\frac{568}{27}  C_F e_u^2  \mathcal{Q}_{f}  
+ \bigg(
        \frac{148}{9}
        -\frac{64}{3} \zeta_3
\bigg) C_F e_u^2 N \mathcal{Q}_{q} \,.
%
\\
 {Z_{ns,f}^{(0,3)}} &=
\bigg(
        -\frac{370}{3}
        +96 \zeta_3
\bigg) e_u^6
-\frac{568}{27} e_u^4  \mathcal{Q}_{f}  
+ \bigg(
        \frac{148}{9}
        -\frac{64 \zeta_3}{3}
\bigg) e_u^2  \mathcal{Q}_f^{(2)}  
+\frac{208}{81} e_u^2 ( \mathcal{Q}_f )^2 \,.
%
%
\\
{ Z_{s, \epsilon}^{(3,0)}} &=
 \frac{1}{\ep^2} \bigg\{ 
-\frac{44}{3} C_A C_F n_F T_F
+\frac{16}{3} C_F n_F^2 T_F^2 
 \bigg\}
+\frac{1}{\ep} \bigg\{  
\frac{218}{9} C_A C_F n_F T_F
-12\, C_F^2 n_F T_F
+\frac{8}{9} C_F n_F^2 T_F^2
\bigg\}\,.
\\
 {Z_{s, \epsilon}^{(2,1)}} &=
\frac{1}{\ep} \bigg\{  
-28 \, C_F n_F T_F e_u^2
+16 \,  C_F T_F \mathcal{Q}_q 
\bigg\}\,.
\\
 {Z_{s, \epsilon}^{(1,2)}} &=
\frac{1}{\ep} \bigg\{  
-28 \, C_F e_u^2  \mathcal{Q}_{f} 
+16 \, C_F e_u^2 N \mathcal{Q}_q 
\bigg\}\,.
\\
 {Z_{s, \epsilon}^{(0,3)}} &=  
\frac{1}{\ep^2} \bigg\{  
\frac{16}{3} e_u^2 \big( \mathcal{Q}_f \big)^2 
\bigg\}
%
+ \frac{1}{\ep} \bigg\{ 
-28\, e_u^4  \mathcal{Q}_f
+ 16\, e_u^2  \mathcal{Q}_f^{(2)} 
+ \frac{8}{9} e_u^2 \big( \mathcal{Q}_f \big)^2
\bigg\}\,.
%
%
\\
{ Z_{s,f}^{(3,0)}} &=
 \big(
    - \frac{326}{27}
        +52 \zeta_3
\big) C_A C_F n_F T_F
+ \big(
        23
       -48 \zeta_3
\big) C_F^2 n_F T_F
+\frac{352}{27} C_F n_F^2 T_F^2 \,.
%
%
\\
 {Z_{s,f}^{(2,1)}} &=
-173 \, C_F e_q^2 n_F T_F
+ \bigg(
         196
        -48 \zeta_3
\bigg) C_F T_F \mathcal{Q}_{q} 
%
%
\\
 {Z_{s,f}^{(1,2)}} &=
-173 \, C_F e_u^2  \mathcal{Q}_{f}  
+ \bigg(
         196
        -48 \zeta_3
\bigg) C_F e_u^2 N \mathcal{Q}_{q} \,.
%
\\
 {Z_{s,f}^{(0,3)}} &=
-173 \, e_u^4  \mathcal{Q}_{f}  
+ \big(
         196
        -48  \zeta_3
\big) e_u^2  \mathcal{Q}_f^{(2)} 
+\frac{352}{27} e_u^2 ( \mathcal{Q}_f )^2 \,.
\end{align}
In the following, we present the three-loop constants for the pseudo-scalar current. We note that the constant $Z_{p,\ep}^{(m,n)}$ also contain the Yukawa contributions.
\begin{align}
{Z_{p,\epsilon}^{(3,0)}} &=
\frac{1}{\epsilon^3} \bigg\{ 
- \frac{121}{9} C_A^2 C_F
- \frac{33}{2} C_A C_F^2 
- \frac{9}{2} C_F^3  
+ \frac{88}{9} C_A C_F n_F T_F
+ 6 C_F^2 n_F T_F  
- \frac{16}{9} C_F n_f^2 T_F^2 
\bigg\} 
+\frac{1}{\epsilon^2} \bigg\{ 
-\frac{257}{54} C_A^2 C_F  
\nonumber\\
&
-\frac{215}{12} C_A C_F^2 
+\frac{9}{4} C_F^3 
+\frac{220}{27} C_A C_F n_f T_F 
+\frac{19}{3} C_F^2 n_f T_F
-\frac{88}{27} C_F n_f^2 T_F^2 
\bigg\} 
+\frac{1}{\epsilon} \bigg\{ 
-\frac{599}{108} C_A^2 C_F 
+\frac{3203}{36} C_A C_F^2 
\nonumber\\
&
-\frac{43}{2} C_F^3 
+\left( -\frac{116}{9} + 16 \zeta_3 \right) C_A C_F n_f T_F
+\left( -\frac{214}{9} - 16 \zeta_3 \right) C_F^2 n_f T_F
+\frac{68}{27} C_F n_f^2 T_F^2  
\bigg\} \,.
%
\\
{Z_{p,\epsilon}^{(2,1)}} &= 
\frac{1}{\epsilon^3} \bigg\{
-\frac{33}{2} C_A C_F e_u^2  
-\frac{27}{2} C_F^2 e_u^2  
+ 6 C_F n_f T_F e_u^2  
\bigg\}
+\frac{1}{\epsilon^2} \bigg\{ 
-\frac{215}{12}  C_A C_F e_u^2 
+\frac{27}{4} C_F^2 e_u^2 
+\frac{31}{3} C_F n_f T_F e_u^2 
\nonumber\\&
-4 C_F T_F \mathcal{Q}_q
\bigg\} 
+ \frac{1}{\epsilon} \bigg\{ 
 \frac{3203}{36} C_A C_F e_u^2 
-\frac{129}{2} C_F^2 e_u^2  
-\frac{253}{9} C_F n_f T_F e_u^2 
+ \left( \frac{13}{3} - 16 \zeta_3 \right)  C_F T_F \mathcal{Q}_q 
\bigg\} \,.
%
\\
{Z_{p,\epsilon}^{(1,2)}} &= 
\frac{1}{\epsilon^3} \bigg\{
-\frac{27}{2} C_F e_u^4 
+6 C_F e_u^2  \mathcal{Q}_{f} 
\bigg\}
+\frac{1}{\epsilon^2} \bigg\{ 
 \frac{27}{4} C_F e_u^4 
+\frac{31}{3} C_F e_u^2 \mathcal{Q}_f
-4 C_F e_u^2 N \mathcal{Q}_q  
\bigg\} 
\nonumber\\&
+ \frac{1}{\epsilon} \bigg\{ 
-\frac{129}{2} C_F e_u^4 
-\frac{253}{9} C_F e_u^2 \mathcal{Q}_f  
+\left( \frac{13}{3} - 16 \zeta_3 \right) C_F e_u^2 N \mathcal{Q}_q  
\bigg\} \,.
%
\\
{Z_{p,\epsilon}^{(0,3)}} &= 
\frac{1}{\epsilon^3} \bigg\{ 
-\frac{9}{2} e_u^6 
+6 e_u^4 \mathcal{Q}_f
-\frac{16}{9} e_u^2 \big(\mathcal{Q}_f\big)^2  
\bigg\} 
+\frac{1}{\epsilon^2} \bigg\{ 
 \frac{9}{4} e_u^6 
+\frac{31}{3} e_u^4 \mathcal{Q}_f 
-4 e_u^2 \mathcal{Q}_f^{(2)} 
\nonumber
-\frac{88}{27} e_u^2 \big(\mathcal{Q}_f\big)^2  
\bigg\} \\&
+\frac{1}{\epsilon} \bigg\{ 
-\frac{43}{2} e_u^6 
-\frac{253}{9} e_u^4 \mathcal{Q}_f 
+\left( \frac{13}{3} - 16 \zeta_3 \right) e_u^2 \mathcal{Q}_f^{(2)}  
+\frac{68}{27} e_u^2 \big(\mathcal{Q}_f\big)^2  
\bigg\}\,.
\\
%
%
%
{Z_{p,f}^{(3,0)}} &= 
 \left( -\frac{958}{27} - 208 \zeta_3 \right) C_A^2 C_F  
+\left( -\frac{800}{27} + 608 \zeta_3 \right) C_A C_F^2  
+\left( \frac{304}{3} - 384 \zeta_3 \right) C_F^3  
+\left( \frac{1712}{81} + \frac{128}{3} \zeta_3 \right) C_A C_F n_f T_F
\nonumber\\&
+\left( -\frac{1160}{27} - \frac{128}{3} \zeta_3 \right) C_F^2 n_f T_F  
+\frac{416}{81}  C_F n_f^2 T_F^2 \,.
%
\\
{Z_{p,f}^{(2,1)}}&= 
 \left( -\frac{800}{27} + 608 \zeta_3 \right)  C_A C_F e_u^2  
+ \left( 304 - 1152 \zeta_3 \right) C_F^2 e_u^2 
- \frac{2048}{27}  C_F n_f T_F e_u^2 
+ \left( \frac{296}{9} - \frac{128}{3} \zeta_3 \right) C_F T_F \mathcal{Q}_q \,.
%
\\
{Z_{p,f}^{(1,2)} }&= 
  \left( 304 - 1152 \zeta_3 \right)   C_F e_u^4
-\frac{2048}{27}  C_F e_u^2 \mathcal{Q}_f
+ \left( \frac{296}{9} - \frac{128}{3} \zeta_3 \right)  C_F e_u^2 N\mathcal{Q}_q \,.
%
\\
{Z_{p,f}^{(0,3)}}&= 
  \left( \frac{304}{3} - 384 \zeta_3 \right)  e_u^6
-\frac{2048}{27}  e_u^4 \mathcal{Q}_f 
+ \left( \frac{296}{9} - \frac{128}{3} \zeta_3 \right) e_u^2 \mathcal{Q}_f^{(2)} 
+  \frac{416}{81} e_u^2 \mathcal{Q}_f^2 \,. 
\end{align}
The finite remainders of the pure-singlet contributions
to the three-loop corrections for the quark axial-vector
form factors are presented below. For completeness, the pure QCD results are also provided.
\begin{align}
 \mathcal{F}_{s,f}^{(3,0)} &=
  \left( 409 - 64 \zeta_2 - \frac{192}{5} \zeta_2^2 \right) C_F^2 n_F T_F
+ \left( -\frac{7403}{9} + \frac{1540}{9}\zeta_2 - \frac{68}{5} \zeta_2^2 + 124 \zeta_3 \right) C_F C_A n_F T_F
\nonumber\\&
+ \left( \frac{1876}{9} - \frac{320 }{9}\zeta_2 \right) C_F n_F^2 T_F^2 \,.
\\
 \mathcal{F}_{s,f}^{(2,1)} &=
  \left( 409 - 96 \zeta_2 - \frac{192}{5} \zeta_2^2 \right) C_F n_F T_F e_u^2
+ 32 \zeta_2 C_F T_F \mathcal{Q}_q  \,.
\\
 \mathcal{F}_{s,f}^{(1,2)} &=
  \left( 409 - 96 \zeta_2 - \frac{192}{5} \zeta_2^2 \right) C_F e_u^2 \mathcal{Q}_f
+ 32 \zeta_2 C_F e_u^2 N \mathcal{Q}_q \,.
\\    
 \mathcal{F}_{s,f}^{(0,3)} &=
  \left( 409 - 96\zeta_2 - \frac{192}{5} \zeta_2^2 \right) e_u^4 \mathcal{Q}_f
+ 32 \zeta_2 e_u^2 \mathcal{Q}_f^{(2)} 
+ \left( \frac{1876}{9} - \frac{320}{9} \zeta_2 \right) e_u^2 (\mathcal{Q}_f)^2 \,.
\end{align}
\end{widetext}

\bibliographystyle{apsrev4-1} 
\bibliography{main}

\end{document}